\documentstyle[multicol,prl,aps,epsf,psfig,epsfig]{revtex}

\begin {document}

\title
{
A Percolation Model of Diagenesis                                                    
}
\author
{
S. S. Manna$^1$, T. Datta$^2$, R. Karmakar$^1$ and S. Tarafdar$^3$
}
\address
{
$^1$Satyendra Nath Bose National Centre for Basic Sciences
    Block-JD, Sector-III, Salt Lake, Kolkata-700098, India \\
$^2$Department of Physics, St. Xavier's College, 30 Park Street, Kolkata 700016, India \\
$^3$Condensed Matter Physics Research Centre, Department of Physics, 
Jadavpur University, Kolkata 700032, India \\
}
\maketitle
\begin{abstract}

   The restructuring process of diagenesis in the sedimentary rocks
is studied using a percolation type model. The cementation and 
dissolution processes are modeled by the culling of occupied
sites in rarefied and growth of vacant sites in dense environments.
Starting from sub-critical states of ordinary percolation the system
evolves under the diagenetic rules to critical percolation configurations.
Our numerical simulation results in two dimensions indicate that the
stable configuration has the same critical behaviour as the ordinary
percolation.


\end{abstract}

\begin{multicols}{2}

\section {Introduction}

   Rocks in general, particularly sedimentary rocks e.g. sandstones, limestones etc.,
have porous structures. Typically such a pore space is a highly branched and inter-connected network.
Study of the pore structure of sedimentary rocks is important
from a practical point of view, in problems such as oil-exploration, ground
water flow, spread of pollutants etc. 

   An interesting property of these rocks is that they appear not to have a finite percolation
threshold \cite {Stauffer}. These rock materials are conducting when the pore space is filled with
saline water. It has been observed that these rock samples show finite conductivity
even when the porosity is less than one percent. This implies that a 
connected network of pores exists in the macroscopic length scale, even when the
porosity i.e., the volume fraction of the void space is very little.

   Several empirical laws reflect this property. Archie's law \cite {Archie} connects the conductivity
$\sigma(\phi)$ and the porosity $\phi$ in the following way:
\begin {equation}
\sigma(\phi) / \sigma_w = a \phi^z
\end{equation}
Here, $\sigma_w$ is the conductivity of water, $a \sim 1$ is an empirical parameter and 
$z \sim 2$ is a non-universal exponent that depends on characteristics of the
rock structure. This law suggests that a finite conductivity exists even in
the limit of $\phi \rightarrow 0$ and therefore the percolation threshold is
zero.

   The permeability $K(\phi)$ of the rock structure is related to the porosity $\phi$
through a similar power law, known as Kozeny equation \cite {Kozeny}:
\begin {equation}
K(\phi) = c \phi^{z'}/S_o^2
\end{equation}
where, $z' \approx 3$, $S_o$ is the specific surface area and $c$ is an empirical constant.
This equation also suggests the global connectivity of the pore space is maintained in the
$\phi \rightarrow 0$ limit. 

   A physical process which is responsible for achieving a connected pore structure
at very low porosities is known as ``diagenesis''. Diagenesis is a complex restructuring process by which
granular systems evolve in geological time scales from unconsolidated, 
high-porosity packings toward more consolidated, less porous structures.
Formation of
sedimentary rocks starts with deposition of sand grains under water or in air
\cite {Phillips,Pett,Sahimi}. Initially this gives an unconsolidated and highly porous
$\sim 40-50\%$ sediment. Sedimentation is followed by compaction under pressure and 
diagenesis, before the consolidated sandstone is formed from
the loosely packed sediment \cite {Chil}. Diagenesis may reduce porosity by an order of
magnitude and permeability by as much as four orders of magnitude \cite {Phillips}.

   The final characteristic of the pore network depends strongly on the
diagenetic process. Sandstones are usually formed under water, which contains
dissolved salts. Depending on the nature of the pore-filling fluids,
salts may be deposited as crystallites in the crevices or along walls of
the rock structure, a process called ``cementation''. Otherwise, portions of the
existing solid structure may get eroded or dissolved out in a ``dissolution''
process. The former decreases the porosity of the rock while the latter
increases porosity. The two processes may take place simultaneously. The
details of the chemical nature of the solid and pore filling fluid
determines whether diagenesis leads finally to a stable structure, or to
a continuously developing structure eventually giving rise to caverns
of macroscopic size.

   Sahimi had classified the theoretical studies of modeling
diagenesis in two ways \cite {Sahimi}. The approach of ``chemical modeling''
relies on solving continuum equations of transport and reactions ignoring
the morphology of the pore space. The second approach is ``geometrical modeling''
in which the reaction kinetics and mass transfer are ignored. These models
start with geometrical descriptions of initial unconsolidated pore space
which evolves under simple rules leading to reduction of porosities but 
maintaining the connectivity. For example the model of Wong et. al. \cite {Wong}
starts with a regular lattice in which each bond is a fluid filled cylindrical
tube of uniform radius and conductivity. This system evolves to a random 
resistor network through a random bond-shrinkage mechanism where randomly 
selected bonds of the network shrinks its radius by a constant factor. This
model maintains global connectivity even in the limit of $\phi \rightarrow 0$
and reproduces power law behaviour as in Archie's law. A second model of
Roberts and Schwartz \cite {Roberts} starts with a Bernal distribution of
dense random spheres of equal radii modeling grains. These spheres grow in unison and
the pore space, i.e. the space not covered by the spheres shrinks its volume.
This model gives a low but non-zero percolation threshold $\phi_c \approx 3.5 \%$.
The bimodal ballistic deposition model (BBDM) \cite {Tarafdar} tries to represent
the deposition realistically, but does not address the problem of diagenesis.

%---------------------------------------------------------------------------
%\begin{center}
\begin{figure}[top]
\centerline{\epsfig{file=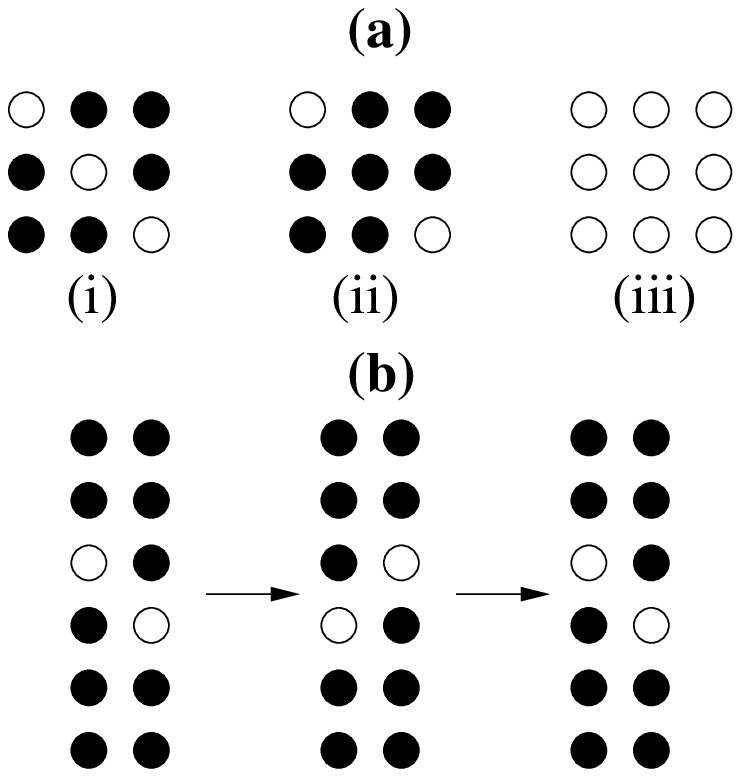, width=5.0 cm}}
\vskip 0.3 cm
\caption{
(a) On an initial configuration as in (i) if the central site is first updated
one gets the SC in (ii). However if the sequential updating rule (I) (see text)
is used one gets the SC as in (iii). (b) In parallel updating this configuration is
locked in a period of cycle 2.
}
\end{figure}
%\end{center}
%---------------------------------------------------------------------------

   In our model, we do not take into 
account the effect of chemical reactions explicitly, so this is also a geometrical 
modeling of diagenesis. We try to simulate the restructuring 
as it may actually occur in porous rocks due to fluid flow. Isolated projected
grains on a wall are smoothed out modeling dissolution, and a gap or cul-de-sac in a solid
is filled by deposition modeling cementation. The restructuring involves two processes. Growth of 
the solid phase at sites with a relatively larger number of occupied nearest 
neighbours, to represent cementation, and removal or culling of occupied sites 
which are isolated, or have too few nearest neighbours, to represent dissolution. 
This algorithm is a stabilizing process leading to a stable structure after several 
time steps. It may be regarded as a self-organizing process as discussed recently 
by several authors \cite {Salmon,Bernabe}.

   Before proposing our model, we briefly describe two other
physical situations which are closely related to our study on diagenesis.
In the Bootstrap percolation model (BPM) occupied sites of a randomly 
occupied regular lattice with certain probability having fewer than certain number of occupied 
neighbours are successively removed \cite {Pollak,Kogut,Chalupa,Adler,Manna,Dhar}. 
On repeated application of this process the system reaches a stable configuration 
where no further sites can be culled. The threshold value of the probability at which the
stable configuration is percolating is calculated.

%---------------------------------------------------------------------------
\begin{center}
\begin{figure}[t]
\centerline{\epsfig{file=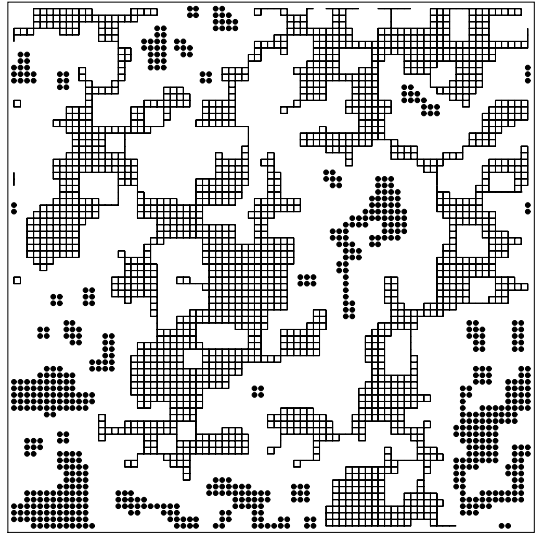, width=7.0 cm}}
\vskip 0.3 cm
\caption{
A stable configuration of the diagenetic percolation for a system of size $L=80$ and 
with $m$=2. Sites on the ``infinite'' incipient cluster are joined by lines and sites 
on the isolated clusters are shown by filled circles.
}
\end{figure}
\end{center}
%---------------------------------------------------------------------------

   Secondly, consider the nearest neighbour Ising model at the zero temperature 
with Glauber spin-flip dynamics in the absence of an external magnetic field. 
Here the direction of a spin follows the direction of the majority of the
neighbouring spins. In the case when there are equal number of up 
and down spins in the neighbourhood, a spin decides its direction with equal probability.
Recently it has been observed in \cite {Spirin} that starting from an arbitrary 
random initial configuration of spins this system 
does not reach the global ground states where all spins are either up or down but
arrive at a frozen two-stripe state in a finite fraction of cases \cite {Spirin}.

   In next section we describe our model and also the updating rules used.
Section III describes our results and we summarize in section IV.

\section {Model}

   In our model, the sites of a regular lattice are randomly occupied $(s_i=1)$ 
with a probability $p$ representing pores and are kept vacant $(s_i=0)$ with a 
probability $1-p$ representing solid grains. This configuration therefore models
the initial unconsolidated porous structure with porosity $\phi(p) = p$.

%---------------------------------------------------------------------------
\begin{center}
\begin{figure}[t]
\centerline{\epsfig{file=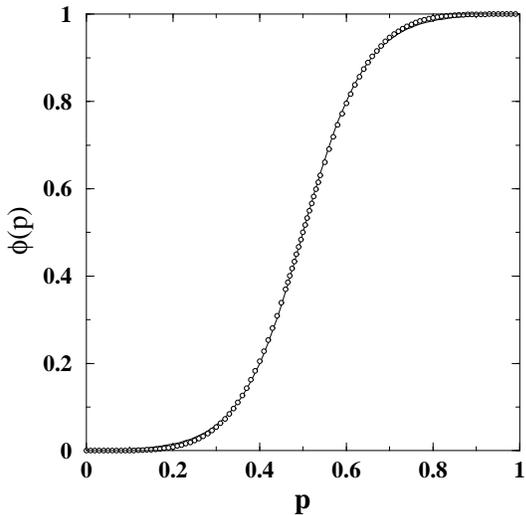, width=7.0 cm}}
\caption{
The porosity $\phi(p)$ as a function of the initial occupation probability $p$ for a system size $L=64$.
The continuous curve is a fit to the data having the form given in Eqn. 1.
}
\end{figure}
\end{center}
%---------------------------------------------------------------------------

   The occupation status of a site $i$ depends on its neighbour number
i.e., the number of occupied neighbours $n_i = \Sigma_j s^i_j$ where, $s^i_j$ is the occupation
of the $j$-th neighbour of the site $i$. All sites of the lattice are
sequentially updated according to the following 
diagenetic conditions: (i) {\it Culling condition}: Occupied sites having fewer than $m$ 
occupied nearest neighbours are vacated i.e., $s_i \rightarrow 0$ if 
$n_i < m$ (ii) the sites with exactly $m$ occupied neighbours remain unaltered
i.e., $s_i \rightarrow s_i$ if $n_i=m$
and (iii) {\it Growing condition}: Vacant sites having more than $m$ occupied
nearest neighbours are occupied i.e., $s_i \rightarrow 1$ if $n_i > m$.

   Starting from the initial configuration the system evolves in different time steps
following these rules. One time step consists of update attempts of all the lattice sites.
One sweep of the lattice results in another occupied configuration which is again 
updated by the same rules. 
This process is continued till the system reaches a stable configuration (SC) 
where no further site changes its occupied or vacant status. In general the SC may have many 
clusters of occupied sites. However, there exists a percolation threshold $p_{mc}$ of $p$
depending on the value of $m$ so that the SC must have a spanning (``infinite'') cluster 
of occupied sites for $p > p_{mc}$ in an infinitely large system.

   Like the cellular automata models, the updating of the sites is important
in our problem. Three possible sequential updating procedures are as follows:
(I) Sites are labeled from 1 to $L^2$ from left to right along a row and from the first row to the last row. 
(II) Only sites with $n_i \ne m$ are randomly selected and updated.
(III) The lattice is divided into odd and even sub-lattices and are updated alternately
but sequentially as in (I). 

%---------------------------------------------------------------------------
\begin{center}
\begin{figure}[t]
\centerline{\epsfig{file=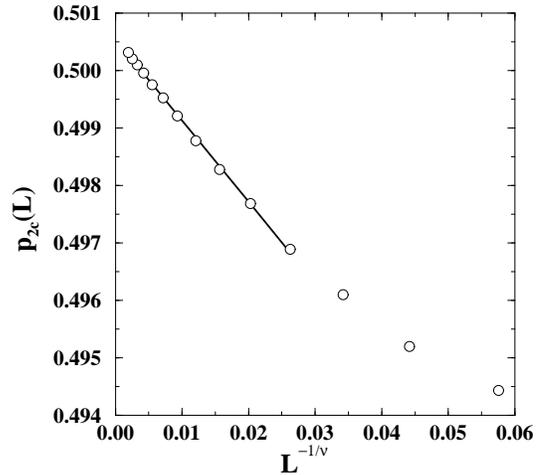, width=7.0 cm}}
\caption{
The plot of the percolation thresholds $p_{2c}(L)$ for $m=2$ for finite system
sizes $L$ as a function of $L^{-1/\nu}$. Using $\nu=4/3$, the correlation length
exponent for ordinary percolation, we get the linear fit for large $L$ values.
The extrapolated value for $p_{2c}$ is 0.5005.
}
\end{figure}
\end{center}
%---------------------------------------------------------------------------

   For BPM, it has been shown in \cite {Manna} that the
SC is independent of the updating sequence. In contrast here the SC does depend
on the updating sequence because culling at one site may inhibit growth at a 
neigbouring site and vice versa. This is seen by considering the neighbour numbers
at all sites of the lattice. When $n_i<m$ the culling of the site $i$ reduces
the neighbour numbers at all neighbouring sites by one i.e., $n_j \rightarrow n_j-1$
where as the growth at $i$ enhances the neighbour numbers at all neighbouring sites 
i.e., $n_j \rightarrow n_j+1$. Therefore the culling at one site may suppress 
the growth at a neighbouring site and vice versa.
In Fig. 1(a) we show an example where two different updating sequences lead to
different SCs. On the other hand, in a fully parallel update all 
sites of the lattice are updated simultaneously at a certain time depending on 
the configuration at the previous time. There may arise some situations as
shown in Fig. 1(b) where a particular cluster of sites never goes to a stable
configuration but takes two different configurations alternately in a
two cycle periodic state.

   A cluster of occupied sites, in which every site has at least $m$ occupied neighbours, is
called an $m$-cluster \cite {Kogut}. Imposition of our diagenetic rules imply that
the surviving clusters in SC must be $m$ clusters. An isolated cluster of 
occupied sites in a $d$-dimensional hypercubic lattice always has some convex
corners on the surface with $d$ neighbours. Therefore in the case when $m \ge d+1$
these sites are always unstable and therefore,
the SC cannot have any finite cluster and has only one infinite $m$-cluster.
A SC at the percolation threshold for $m=2$ is shown in Fig. 2, the smallest
isolated clusters being of size 4.

%---------------------------------------------------------------------------
\begin{center}
\begin{figure}[t]
\centerline{\epsfig{file=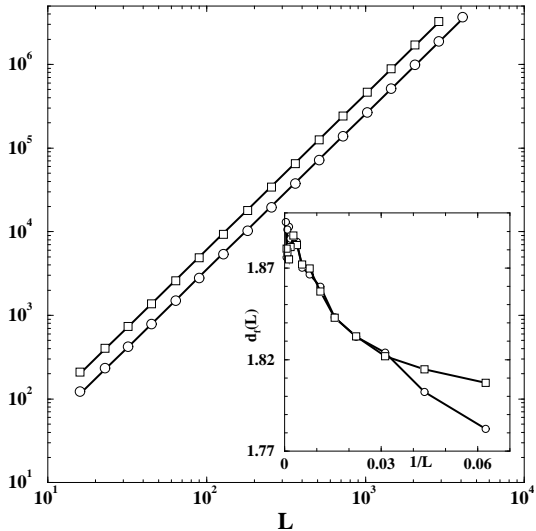, width=7.0 cm}}
\caption{
The average mass of the infinite cluster $S^1_{\infty}$ (circle) and the largest
cluster $2 S^2_{\infty}$ (square) at the diagenetic percolation threshold $p_{2c}(L)$
are plotted with the system size $L$. Average fractal dimension is obtained from
the linear fits shown by continuous lines. The inset shows the variations of the local
slopes $d_f(L)$ and we conclude a value of the fractal dimension $d_f = 1.89 \pm 0.02$.
}
\end{figure}
\end{center}
%---------------------------------------------------------------------------

\section {Results}

   The average fraction $\phi(p)$ of the occupied sites in the SC is defined as
the porosity of this model. We measure this porosity as a function of the
probability $p$ and this variation is plotted in Fig. 3. We find no trace of any
system size dependence on this variation. A functional form like 
\begin {equation}
\phi(p) = 1/[1+\exp((1/2-p)/\Delta p)]
\end {equation}
fits very well to this data with a value of $\Delta p=$ 0.072. The data as well as
the fit are very well consistent to $\phi(1/2)=1/2$ as expected from
the symmetry of occupied and vacant sites.
Compared to the porosity $\phi(p)=p$ in the initial random distribution of occupied 
and vacant sites, the porosity in SC is reduced by an order of magnitude when
$p < 0.35$. We consider this as the reflection of the diagenesis process
in nature observed in our model.

   Like BPM, the culling condition in our model does not
contribute to change the percolation threshold for $m \le 2$. For example, nothing
is culled for $m$=0, isolated sites are culled for $m=1$ and the dangling chain of sites
are culled for $m=2$. Since the connectivity of the system is not affected by these
culling processes, the difference between $p_{mc}$ and $p_c(ord)$ is due to the growing
condition for $m \le 2$, where $p_c(ord)$ is the ordinary percolation threshold. 
For $p$ values very close to $p_c(ord)$ but smaller than it, there may be some 
initial configurations which are not connected because of the presence of only
few vacant sites. If these sites have more than $m$ occupied neighbours they will now be occupied
ensuring the global connectivity of the system. Therefore it is expected that the
$p_{mc} \le p_c(ord)$ for $m \le 2$ on an arbitrary lattice. 
Therefore as the growth rule helps in attaining a connectivity in the system
we expect $p_{mc} \le p_c(BPM)$
for any arbitrary lattice and for any arbitrary value of $m$.
%For the square lattice
%with coordination number $z=4$, $n_i$ can have five possible values from 0 to 4 and
%$m=2$ is the symmetric middle number of these 5 values. Since in this problem
%there is a symmetry between the occupied and vacant sites, we expect that
%$p_{mc} = p_{(z-m)c}$.

%---------------------------------------------------------------------------
\begin{center}
\begin{figure}[t]
\centerline{\epsfig{file=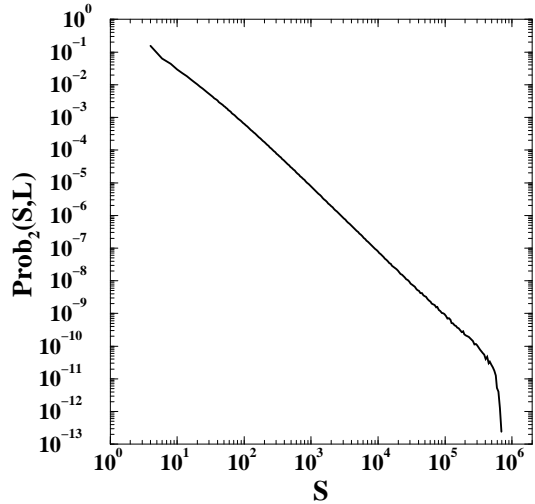, width=7.0 cm}}
\caption{
The distribution of the cluster sizes of the stable configuration at the percolation 
threshold of a lattice of size $L=2048$ and with $m=2$.
}
\end{figure}
\end{center}
%---------------------------------------------------------------------------

   The percolation threshold $p_{mc}$ is the minimum value of the probability
$p$ beyond which an infinite cluster of occupied sites exists with probability 
one in the SC on an infinitely large system. However, for systems of finite 
size this threshold $p_{mc}(L)$ depends on the system size. The correlation 
function $g(r)$ for the percolation problem is defined as the probability that 
a site at a distance $r$ apart from an occupied site belongs to the same cluster. 
For $p < p_{mc}$ the correlation function is expected to decay exponentially as 
$g(r) \sim \exp(-r/\xi)$ where the correlation length $\xi$, a measure 
of the typical cluster diameter, diverges as $\xi \sim (p_{mc}-p)^{-\nu}$ where, 
$\nu$ is the correlation length exponent for the diagenetic percolation.

   We use the standard method of estimating the value of the percolation threshold.  
Using a specific sequence of random numbers, the lattice is filled at some high 
value of $p=p_{hi}$ such that its SC has an infinite cluster. Similarly using
the same sequence of random numbers, the lattice is filled at some low value of
$p=p_{lo}$ so that its corresponding SC does not have an infinite cluster. It is
then similarly tried at a $p=(p_{hi}+p_{lo})/2$. If its SC is connecting then
$p_{hi}$ is equated to $p$, otherwise $p_{lo}$ is equated to $p$. This process
is continued till the difference $(p_{hi}-p_{lo})$ is less than a certain pre-assigned
small number $\epsilon = 10^{-5}$ when $p(seq)=(p_{hi}+p_{lo})/2$ is taken for the percolation
threshold for this particular sequence of random numbers. 
Averaging over the $p(seq)$ values for a large number of independent random number sequences
one obtains the estimate for $p_{mc}(L)$.

%---------------------------------------------------------------------------
\begin{center}
\begin{figure}[t]
\centerline{\epsfig{file=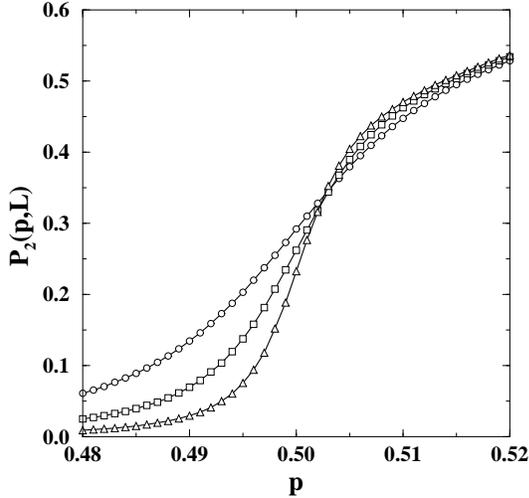, width=7.0 cm}}
\caption{
The plot of the percolation probability $P_2(p,L)$ for $m=2$ of the diagenetic percolation  
for three different system sizes $L$ = 256 (circle), 512 (square) and 1024 (triangle).
}
\end{figure}
\end{center}
%---------------------------------------------------------------------------

   In this process, we tune the probability $p$ to the percolation threshold $p_{mc}(L)$
on a system of size $L$ so that the correlation length is of the same order as
the system size. Therefore for $m=2$, $L \sim (p_{2c}-p_{2c}(L))^{-\nu}$ which implies 
\begin {equation}
p_{2c}(L) = p_{2c}+A.L^{-1/\nu}
\end {equation}
We plot $p_{2c}(L)$ in Fig. 4 with $L^{-1/\nu}$
and we try $\nu=4/3$, the value for the correlation length exponent in the ordinary
percolation. We observe a linear variation for large $L$ values.
On extrapolation, we find a slightly larger value
of $p_{2c}= 0.5005(2)$ for sequential updating of type (I) as stated above.
The $p_{mc}(L)$ values for $L$=2048, 2896 and 4096 are found larger
than 1/2. For the random and sub-lattice sequential updatings the $p_{2c}$ values
are 0.5013 and 0.5009 respectively.
These values should compared to the ordinary percolation threshold of 0.592746
on square lattice \cite {Ziff}.

   Since $m=2$ is the middle point of the five possible values of the neighbour
numbers on a square lattice (i.e. from 0 to 4) and due to the equivalence of vacant
and occupied sites, it may be expected that $p_{2c}$ should be exactly equal to
1/2. However, we argue that the value of $p_{2c}$ very close to 1/2 is actually
accidental and there is no reason why it should be 1/2. We believe that first
appearnce of the global connectivity through occupied sites determining the 
percolation threshold is a very special situation and since we want this 
connectivity through the occupied sites we break the symmetry between the occupied
and vacant sites. An estimate of the similar percolation threshold for $m=3$
on a simple cubic lattice gives a value much lower than 1/2. However, for $m=3$ on the
triangular lattice the value of the percolation threshold is obtained as $0.50 \pm 0.01$.

%---------------------------------------------------------------------------
\begin{center}
\begin{figure}[t]
\centerline{\epsfig{file=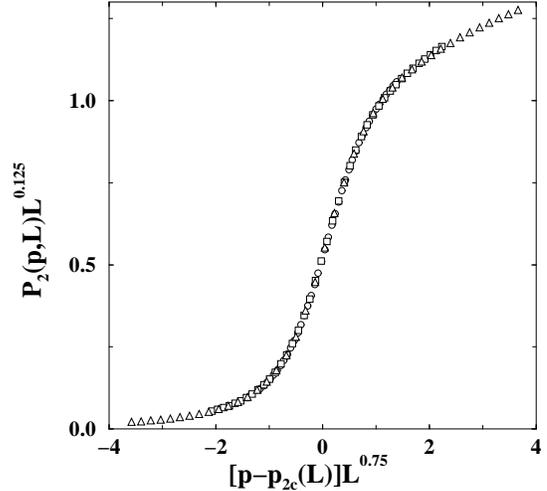, width=7.0 cm}}
\caption{
The diagenetic percolation probability $P_2(p,L)$ as shown in
the previous figure for different system sizes are scaled as
$P_2(p,L)L^{\beta/\nu}$ with $[p-p_{2c}(L)]L^{1/\nu}$. The collapse
is obtained using $\nu$=4/3 and $\beta/\nu$ =0.125.
}
\end{figure}
\end{center}
%---------------------------------------------------------------------------

   The $p(seq)$ values corresponding to different sequences of random numbers are spread 
around their mean value $p_{mc}(L)$. The root mean square deviation from the average value
\begin {equation}
\Delta(L) = (<p(seq)^2>-[p_{mc}(L)]^2)^{1/2}
\end {equation}
is supposed to have a dependence on the system size
$L$ as: $\Delta (L) \sim L^{-1/\nu}$.
Plotting  $\Delta(L)$ vs. $L$ for $m=2$ on a double
logarithmic scale which fits nicely to a straight line gives
a value for the correlation length exponent $\nu = 1.35(2)$. 

   The fractal dimension $d_f$ of the ``infinite'' incipient cluster (IIC) 
of the SC exactly at the percolation threshold is also calculated. A large 
number of SCs are generated at $p=p_{2c}$. The average size 
$S_\infty$ of the IIC is calculated in two ways: (i) Average size 
$S^1_\infty(L)$ of the infinite clusters is measured over the
spanning SCs only (ii) Average size $S^2_\infty(L)$ of the largest cluster 
is calculated over all SCs. Using the definition of percolation probability
as defined below, $S^2_\infty(L) = L^2 P_2(p_{2c}(L),L)$.
Both measures of the IIC are expected to give 
the fractal dimension: $S^{1/2}_\infty(L) \sim L^{d_f}$. In Fig. 5 we plot 
both $S^1_\infty(L)$ and $2 S^2_\infty(L)$ with L on a 
double logarithmic scale for the system sizes varying from 16 to 4096.
The average slopes are 1.863 and 1.860 for $S^1_\infty(L)$ and $S^2_\infty(L)$
respectively. Further, we plot the local slopes $d_f(L)$ with $1/L$ in the
inset of Fig. 5. After considerable variation over the small systems
the fractal dimension seems to converge at $1.89 \pm 0.02$ for the large
system sizes compared to 91/48 of the ordinary percolation \cite {Stauffer}.

%---------------------------------------------------------------------------
\begin{center}
\begin{figure}[t]
\centerline{\epsfig{file=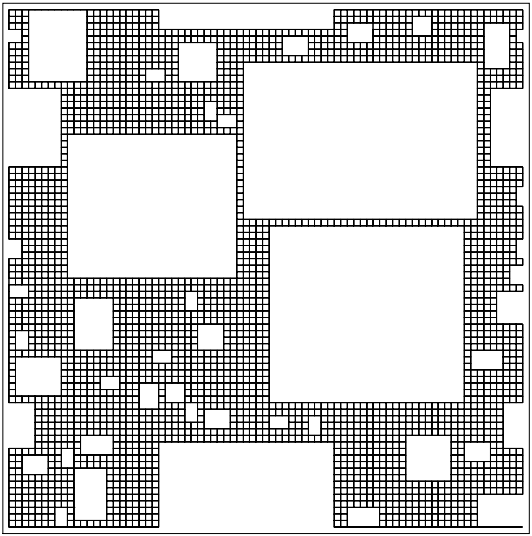, width=7.0 cm}}
\vskip 0.3 cm
\caption{
A stable configuration (SC) at the percolation threshold for a system 
size of $L=80$ and with $m$=3. Sites on the ``infinite'' incipient cluster are
joined by lines.
}
\end{figure}
\end{center}
%---------------------------------------------------------------------------

   The cluster size distribution of occupied sites on the SCs are 
also measured at the percolation threshold. We define Prob$_2(S,L)$ 
as the probability of a cluster of $S$ occupied sites on a SC of a 
system of size $L$ with $m=2$. We start from many independent 
configurations at $p_{2c}(L) \approx 0.5001$ for $L=2048$.
These are sub-critical configurations for the ordinary percolation.
We measure the Prob$_2(S,L)$ at each time step and keep track of how
this distribution changes from the initial exponential distribution to
the power law distribution as shown in Fig. 6. We notice that at very
short times of the order of 1, the distribution takes the form of the
steady state distribution. In this distribution we do not include the 
``infinite'' cluster spanning the system. As expected the distribution 
appears to be a power law: Prob$_2(S,L) \sim S^{-\tau}$ where
$\tau = 2.02 \pm 0.06$ is obtained compared to 187/91 for the ordinary
percolation \cite{Stauffer}.

%---------------------------------------------------------------------------
\begin{center}
\begin{figure}[t]
\centerline{\epsfig{file=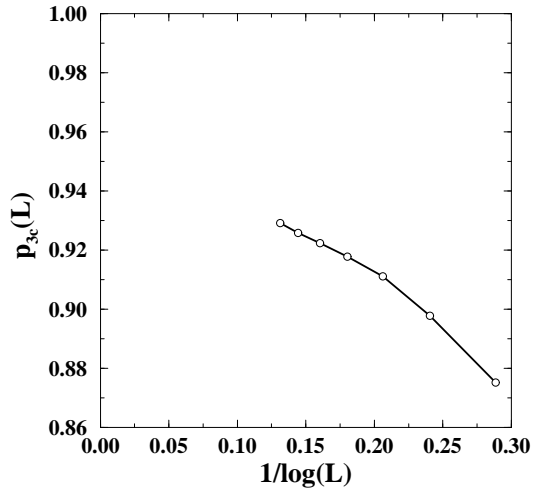, width=7.0 cm}}
\vskip 0.3 cm
\caption{
Plot of $p_{3c}(L)$ with $1/\log(L)$ which on extrapolation to
$L \rightarrow \infty$ gives $p_{3c}=0.96 \pm 0.01$.
}
\end{figure}
\end{center}
%---------------------------------------------------------------------------

   The order parameter is the percolation probability $P_m(p)$
that is the average fraction of sites on the largest
occupied cluster in the SC. For finite systems it is denoted
by $P_m(p,L)$. Variation of the percolation probability is shown
in Fig. 7 and it varies as:
\begin {equation}
P_m(p,L) \sim [p-p_{mc}(L)]^\beta
\end {equation}
This variation is true in the limit of $L \rightarrow \infty$.
For finite systems however, according to the scaling theory
\cite {Stauffer}, the scaling variable should be $L/\xi$, where
the correlation length is defined as $\xi = [p-p_{mc}(L)]^{-\nu}$.
Therefore for a finite system of size $L$ the variation of
percolation probability should be:
\begin {equation}
P_m(p,L) = L^{-\beta/\nu}F[(p-p_{mc}(L))L^{1/\nu}]
\end {equation}
where, the scaling function $F(x) \rightarrow x^{\beta}$ for large $L$.
We show the collapse of the data in Fig. 8 using this scaling formulation.
We again try $\nu=4/3$ and then obtain a value of $\beta/\nu = 0.125$
for the data collapse, giving $\beta = 0.166$ compared to 5/36 for the
ordinary percolation \cite {Stauffer}.

   Next we studied the case of $m$=3 on the square lattice. In this case the SC 
can only be completely vacant or it can have only one infinite cluster but cannot have 
isolated clusters. Since in general 
there will always be some sites which have less than 3 occupied neighbours
on the surface of an isolated cluster, these sites will be unstable under
the diagenetic rules and the cluster will therefore cannot survive in SC.
In Fig. 9 we show the picture of a SC for $m$=3.
It is a simple spanning cluster having many rectangular holes as in BPM \cite {Adler}.
The percolation threshold $p_{3c}(L)$ also has $L$ dependence and on extrapolation
with $1/\log(L)$ (as was done in BPM) we get $p_{3c}=0.96 \pm 0.01$ (Fig. 10).

\section {Summary}

   The restructuring process of diagenesis in sedimentary rocks involving
cementation and dissolution has been studied by a percolation model.
Simulations on a square lattice shows that the porosity is highly reduced
due to restructuring as is observed in rocks. We also observe that 
starting from the sub-critical configurations of ordinary percolation at a 
certain threshold value $p_{2c}$ of the pore probability the system evolves 
to a globally connected porous space at the stable state. This configuration
is critical since it shows long range correlations. Our numerical results
give strong indications that the stable states in this model have the same 
critical behaviour as that of ordinary percolation. We view the dynamics
under diagenetic rules as a self-organizing dynamics in a limited sense
since one has to tune $p$ to arrive at a specific sub-critical configuration
at $p_{2c}$ so that it organizes to show criticality in the stable state.

   After completion of a draft of this manuscript we were informed by the 
editor that a spin model on a square lattice where each spin $\pm 1$ 
was flipped only when more than half of its four neighbours point into 
the opposite direction was studied in \cite {Stauffer97}. Using a much
bigger system size $(L \approx 7 \times 10^5)$ compared to what we used
a percolation threshold of $0.5007 \pm 0.0001$ was estimated which is
consistent with our results.

\leftline {Electronic Address: manna@boson.bose.res.in}

\end{multicols}
\end {document}